\documentclass[12pt]{article}
\usepackage{amsmath,amssymb}
\begin{document}
\title{Black holes and the quark-gluon plasma\footnote{Research supported in part by the US Department of Energy under grant DE-FG05-91ER40627.}}
\author{George Siopsis\footnote{E-mail: siopsis@tennessee.edu}\\
\em Department of Physics
and Astronomy, \\
\em The University of Tennessee, Knoxville, \\
\em TN 37996 - 1200, USA.
}
\date{}
                                                                                
\maketitle
\vspace{-3.5in}\hfill UTHET-09-0101\vspace{3.5in}
                                                                                
\abstract{I discuss the possibility that the quark-gluon plasma at strong coupling admits a description in terms of a black hole in asymptotically anti-de Sitter space.}

\vspace{3in}

\noindent {\it Prepared for the Proceedings of Recent Developments in Gravity -- NEB XIII, Thessaloniki,
Greece, June 2008.}
\newpage

\renewcommand{\a}{\alpha}
\renewcommand{\b}{\beta}
\renewcommand{\c}{\gamma}
\renewcommand{\d}{\delta}
\newcommand{\m}{\mu}
\newcommand{\n}{\nu}
\def\o{\omega}
\def\p{\partial}
\def\t{\tau}
\def\A{\mathcal{A}}
\def\B{\mathcal{B}}
\def\F{\mathcal{F}}
\def\H{\mathcal{H}}
\def\K{\mathcal{K}}
\def\pK{{K}}
\def\N{\mathcal{N}}
\def\O{\mathcal{O}}
\def\W{\mathcal{W}}
\def\2{\frac{1}{2}}
\def\4{\frac{1}{4}}
\def\G{\Gamma}
\def\L{\Lambda}
\newcommand{\llr}{\langle\lambda\rangle}

\def\bea{\begin{eqnarray}}
\def\eea{\end{eqnarray}}
\def\bes{\begin{eqnarray}}
\def\ees{\end{eqnarray}}
\def\beq{\begin{eqnarray}}
\def\eeq{\end{eqnarray}}
\def\be{\begin{equation}}

\def\ee{\end{equation}}

\def\ba{\begin{array}}
\def\ea{\end{array}}
\def\bi{\begin{itemize}}
\def\ei{\end{itemize}}

\def\tr{{\rm tr}}
\def\Tr{{\rm Tr}}

\def\lesssim{\mathrel{\mathpalette\vereq<}}

\def\vereq#1#2{\lower3pt\vbox{\baselineskip1.5pt \lineskip1.5pt

\ialign{$#1\hfill##\hfil$\crcr#2\crcr\sim\crcr}}}

\def\gtrsim{\mathrel{\mathpalette\vereq>}}

\def\dslash{\not{\hbox{\kern-2pt $\partial$}}}

\def\eslash{\not{\hbox{\kern-2pt $\epsilon$}}}

\def\Dslash{\not{\hbox{\kern-4pt $D$}}}

\def\Aslash{\not{\hbox{\kern-4pt $A$}}}

\def\Qslash{\not{\hbox{\kern-4pt $Q$}}}

\def\Wslash{\not{\hbox{\kern-4pt $W$}}}

\def\pslash{\not{\hbox{\kern-2.3pt $p$}}}

\def\kslash{\not{\hbox{\kern-2.3pt $k$}}}

\def\qslash{\not{\hbox{\kern-2.3pt $q$}}}

\def\np#1{{\sl Nucl.~Phys.~\bf B#1}}

\def\pl#1{{\sl Phys.~Lett.~\bf B#1}}

\def\mpl#1{{\sl Mod.~Phys.~Lett.~\bf A#1}}

\def\pr#1{{\sl Phys.~Rev.~\bf D#1}}

\def\prl#1{{\sl Phys.~Rev.~Lett.~\bf #1}}

\def\cpc#1{{\sl Comp.~Phys.~Comm.~\bf #1}}

\def\cqg#1{{\sl Class.~Quant.~Grav.~\bf #1}}

\def\cmp#1{{\sl Commun.~Math.~Phys.~\bf #1}}

\def\anp#1{{\sl Ann.~Phys.~(NY) \bf #1}}

\def\etal{{\em et al.}}

\def\half{{\textstyle\frac{1}{2}}}
\def\halfi{{\textstyle\frac{i}{2}}}
\def\three{{\textstyle\frac{1}{3}}}
\def\four{{\textstyle\frac{1}{4}}}

\def\halfhalf{{\textstyle\frac{1}{4}}}

\def\bi{\begin{itemize}}

\def\ei{\end{itemize}}

\def\tr{{\rm tr}}

\def\Tr{{\rm Tr}}

\section{Introduction}

\vspace{1cm}

{\it
``A second unexpected connection comes from studies carried out using the Relativistic Heavy Ion Collider, a particle accelerator at Brookhaven National Laboratory. This machine smashes together nuclei at high energy to produce a hot, strongly interacting plasma. Physicists have found that some of the properties of this plasma are better modeled (via duality) as a tiny black hole in a space with extra dimensions than as the expected clump of elementary particles in the usual four dimensions of spacetime. The prediction here is again not a sharp one, as the string model works much better than expected. String-theory skeptics could take the point of view that it is just a mathematical spinoff. However, one of the repeated lessons of physics is unity - nature uses a small number of principles in diverse ways. And so the quantum gravity that is manifesting itself in dual form at Brookhaven is likely to be the same one that operates everywhere else in the universe.''}

\rightline{-- Joe Polchinski}

\vspace{1cm}

The AdS/CFT correspondence~\cite{adscft1,adscft2,adscft3,adscftrev} which has emerged from string theory has provided a path for understanding a conformally invariant gauge theory (CFT) at strong coupling in terms of a dual gravitational description in an asymptotically anti-de Sitter (AdS) space in one higher dimension.
The quark-gluon plasma observed at the Relativistic Heavy Ion Collider (RHIC) is thought to be strongly interacting \cite{RHICrev1,RHICrev2,RHICrev3}. Even though it is described by a different gauge theory, namely Quantum Chromodynamics (QCD), one hopes that there are enough similarities between the two theories allowing us to acquire
a good qualitative (if not quantitative) understanding of its behavior.
 This has cultivated considerable interest in the connection between string theory and the experimental results at RHIC \cite{sinzahed,sonstarinets,zahedsin,son1,son2,son3}.

Via the AdS/CFT correspondence, information on the strongly interacting plasma is gained by studying the supergravity solution in AdS space.  From the AdS metric one may then calculate the stress-energy tensor of the dual gauge theory using holographic renormalization \cite{Skenderis1,Skenderis2}.  The simplest supergravity solution to study is the AdS Schwarzschild metric which has been shown to be dual to a static fluid on the AdS boundary.  This analysis has been extended to include small deformations of the Schwarzschild metric governed by quasinormal modes that dictate the late-time behavior of the black hole perturbations \cite{FGMP}.  These modes have been studied in great detail (see \cite{QNM1,QNM2,QNM3,QNM4} and references therein).  The lowest frequency modes govern the hydrodynamic behavior of the conformal field theory on the boundary \cite{PSS1,PSS2,Siopsis}.  

The lowest lying gravitational quasinormal modes for an AdS Schwarzschild black hole with a spherical horizon were calculated numerically in four and five dimensions in \cite{FGMP,MP} and analytically in arbitrary dimension in \cite{Siopsis}. They were shown to be in agreement with appropriate sinusoidal perturbations of the gauge theory plasma on the AdS boundary.  To make contact with experimental results, the boundary is defined through a foliation of flat surfaces turning it into flat Minkowski space. This results in a
finite ``conformal soliton flow'' once the spherical AdS boundary one obtains in global coordinates is conformally mapped to the physically relevant flat Minkowski spacetime \cite{FGMP}.  The perturbations also allow for calculations of the elliptic flow of the plasma and its thermalization time -- both observable at RHIC.  While there is still work to be done, these calculations compare well with what has been found experimentally.

An alternative to a spherical AdS black hole is one with a hyperbolic horizon \cite{mann1,mann2,Vanzo:1997gw,Brill:1997mf,Birmingham,Emparan:1999gf,BS1,BS2}.
They are usually referred to as topological AdS black holes because they possess topologically non-trivial horizons.
Their effect on the gauge theory plasma on the AdS boundary was elucidated in \cite{AS3}.
Gravitational perturbations were shown to possess quasinormal modes whose lifetime was comparable to or longer than their counterparts in the case of spherical horizons.
These results were in agreement with those obtained by studying the hydrodynamics of the gauge theory plasma on the boundary.
Therefore, topological AdS black holes might have a significant effect on the behavior of the quark-gluon plasma in heavy ion collisions at RHIC and the LHC.


The AdS/CFT correspondence has been applied to the description of the quark-gluon plasma in several different scenarios; see \cite{MV,Shuryak} for nice reviews.  Janik and Peschanski showed that a model for relativistic heavy ion collisions, suggested two decades previously by Bjorken \cite{Bjorken}, can be seen to be a consequence of the AdS/CFT correspondence \cite{JP}.  They obtained an asymptotic solution to the Einstein equations in the five-dimensional bulk in the large longitudinal proper time ($\tau$) limit and showed that it gave rise to hydrodynamics on the four-dimensional boundary matching the Bjorken flow of an ideal fluid \cite{Bjorken}.  The work was subsequently continued to include subleading terms in the large $\tau$ expansion which could be understood as dissipative effects in the hydrodynamic behavior of the gauge theory plasma \cite{NakSin,Janik,SinNakKim,ASM}.

Kajantie, {\em et al.,} \cite{KLT} showed that the two-dimensional Bjorken flow could be derived from a static black hole in the three-dimensional bulk spacetime by an appropriate time-dependent slicing near its boundary.
This result was then extended to arbitrary dimensions \cite{AS} by showing that to leading order in $\tau$ there exists a foliation near the boundary of an AdS Schwarzschild black hole which corresponds to Bjorken flow in the dual gauge theory plasma. The Schwarzschild metric to leading order in $\tau$ agreed with the time-dependent asymptotic solution of Janik and Peschanski \cite{JP} in five dimensions. In three dimensions it reduced to the result of Kajantie, {\em et al.,} \cite{KLT}.

Subleading corrections in the large $\tau$ expansion were discussed in \cite{AS2}.
It was shown that next-to-leading-order corrections correspond to viscosity in the gauge theory plasma. At this level the coefficient of viscosity $\eta$ is arbitrary, in agreement with results in five dimensions based on an asymptotic time-dependent solution to the Einstein equations \cite{NakSin}.
At next-to-next-to-leading order the Schwarzschild metric yields a flow which is not boost-invariant no matter how one chooses the foliation near the AdS boundary. Boost invariance is recovered after the Schwarzschild metric is perturbed by a power law $\tau$-dependent perturbation. The perturbed metric was shown to be non-singular in the bulk, provided
\be\label{eq1} \frac{\eta}{s} = \frac{1}{4\pi} \ee
where $s$ is the entropy density, in agreement with asymptotic time-dependent solutions in five dimensions \cite{Janik}.
This special value of the ratio $\eta/s$ is also in agreement with the case of sinusoidal perturbations of an AdS Schwarzschild black hole \cite{son1,son2,son3}.


The outline of my presentation is as follows.
In section \ref{sec2}, I discuss the various types of AdS Schwarzschild black holes and the dual static gauge theory fluid on the AdS boundary.
In section \ref{sec3} I discuss perturbations and the calculation of low frequency quasinormal modes. These modes are then shown to agree with the hydrodynamic behavior of the gauge theory fluid in section \ref{sec4}.
In section \ref{sec5} a different foliation of the AdS Schwarzschild spacetime near the AdS boundary is shown to lead to Bjorken flow for the gauge theory fluid.
Finally section \ref{sec6} contains my concluding remarks.


\newcommand{\qn}{\textswab{q}}
\newcommand{\wn}{\textswab{w}}
\renewcommand{\d}{\partial}
\renewcommand{\O}{\hat{\cal O}}
\newcommand{\q}{\bm{q}}
\newcommand{\x}{\bm{x}}
\newcommand{\gYM}{g_{\mathrm{YM}}}
\def\ofo{ { {}_2 \! F_1 }}
\newcommand{\stru}{\rule[-.2in]{0in}{.2in}}

\section{AdS Schwarzschild black holes}
\label{sec2}


The metric of an AdS Schwarzschild black hole in $d$ dimensions can be written in static coordinates as
\be\label{metric}
ds_{\mathrm{b.h.}}^2=-\left(\frac{r^2}{R^2}+\pK-\frac{2\mu}{r^{d-3}}\right)dt^2+\frac{dr^2}{\frac{r^2}{R^2}+\pK-\frac{2\mu}{r^{d-3}}}+r^2d\Sigma^2_{\pK,d-2}
\ee
For simplicity I shall choose units so that the AdS radius $R=1$.
The horizon radius and Hawking temperature are, respectively,
\be\label{BHT}
2\mu=r_+^{d-1}\left( 1+\frac{\pK}{r_+^2}\right)~,~~~~T_H=\frac{(d-1)r_+^2+\pK(d-3)}{4\pi r_+}
\ee
The mass and entropy of the hole are, respectively,
\be\label{BH}
M=(d-2)(\pK+r_+^2)\frac{r_+^{d-3}}{16\pi G} Vol(\Sigma_{\pK,d-2})~,~~~ S=\frac{r_+^{d-2}}{4G} Vol(\Sigma_{\pK,d-2})
\ee
$\pK =0$ corresponds to a
flat horizon ($\mathbb{R}^{d-2}$),
$\pK =+1$ to a spherical horizon ($\mathbb{S}^{d-2}$) and
$\pK = -1$ to a hyperbolic horizon ($\mathbb{H}^{d-2} /\Gamma$, where $\Gamma$ is a discrete group of isometries of $\mathbb{H}^{d-2}$ (topological black hole)).

The harmonics on $\Sigma_{\pK,d-2}$ obey the equation
\be
\left(\nabla^2 + k^2\right){\mathbb T}=0
\ee
For $\pK = 0$, $k$ represents momentum (continuous parameter),
for $\pK = +1$,
\be
k^2=l(l+d-3)-\delta
\ee
where $l$ is the (discrete) angular momentum quantum number
and for $\pK = -1$,
%
\be
k^2=\xi^2+\left(\frac{d-3}{2}\right)^2+\delta
\ee
where $\xi$ is a discrete parameter for non-trivial $\Gamma$ \cite{HyperB1,HyperB2,HyperB3,HyperB4,HyperB5,HyperB6,HyperB7}.
$\delta =0,1,2$ for scalar, vector, or tensor harmonics, respectively.

To study the consequencies of the AdS/CFT correspondence, it is convenient to
switch coordinates to
\be z = \frac{1}{r} \ee
placing the boundary of AdS space at $z=0$.
For the flat (large) Schwarzschild black hole
($\pK = 0$), the metric (\ref{metric}) reads
\be\label{eqmebh} ds_{\mathrm{b.h.}}^2 = \frac{1}{z^2} \left( -(1-2\mu z^{d-1})  dt^2 + {d\vec x\,}^2 + \frac{dz^2}{1-2\mu z^{d-1}} \right) \ee
where $\vec x \in \mathbb{R}^{d-2}$. The horizon is at
\be\label{eq5} z_+ = (2\mu)^{-\frac{1}{d-1}} \ee
and the
Hawking temperature (\ref{BHT}) reads
\be\label{eqTH} T_H = \frac{d-1}{4\pi z_+} \ee
The above expressions are related to a gauge theory fluid on the boundary via holographic renormalization \cite{Skenderis1,Skenderis2}.
To see this, write the metric (\ref{eqmebh})
in Fefferman-Graham coordinates
\be\label{FG}
ds^2=\frac{g_{\mu\nu}dx^\mu dx^\nu+dz_{FG}^2}{z_{FG}^2}\ee
Then near the
boundary at $z_{FG}=0$ expand
\be
g_{\mu\nu} = g_{\mu\nu}^{(0)}+z_{FG}^2 g_{\mu\nu}^{(2)} +
\dots+z_{FG}^{d-1}g_{\mu\nu}^{(d-1)} +h_{\mu\nu}^{(d-1)}z_{FG}^{d-1}\ln z_{FG}^2 +\mathcal{O}(z_{FG}^d) \ee
where $g^{(0)}_{\mu\nu}=\eta_{\mu\nu}$.

One obtains the stress-energy tensor of the plasma
\be\label{eqgT} \langle T_{\mu\nu}\rangle = \frac{d-1}{16\pi G} g_{\mu\nu}^{(d-1)}
\ee
leading to the energy density and pressure, respectively,
\be\label{eq15} \epsilon = \langle T^{tt} \rangle = (d-2) \frac{\mu}{8\pi G} \ \ , \ \ \ \ p = \langle T^{ii} \rangle = \frac{\mu}{8\pi G} \ee
obeying $p = \frac{1}{d-2} \epsilon$, as expected of a conformal fluid in $d-2$ spatial dimensions.

The equation of state of this fluid is
\be p = \frac{1}{16\pi G_d} \left( \frac{4\pi T_H}{d-1} \right)^{d-1} \ee
We may also write the energy and entropy densities, respectively, as functions of the temperature (\ref{eqTH})
\be\label{eqes} \epsilon = \frac{d-2}{16\pi G} \left( \frac{4\pi T_H}{d-1} \right)^{d-1} \ \ , \ \ s = \frac{dp}{dT} = \frac{1}{4G} \left( \frac{4\pi T_H}{d-1} \right)^{d-2} \ee
It should be noted that we obtained a static fluid.

The above results can be easily generalized to the case of a curved horizon (boundary), i.e., $\pK = \pm 1$.

\section{Perturbations}
\label{sec3}
As was shown by Ishibashi and Kodama \cite{IK}, any perturbation of the black hole background (\ref{metric}) obeys a radial wave equation of the Schr\"odinger-like form (master equation)
\be\label{ME}
-\frac{d^2\phi}{dr_*^2} + V[r(r_*)] \phi = \omega^2 \phi
\ee
where $V(r)$ is a potential whose form depends on the perturbation and
the ``tortoise coordinate'' $r_*$ is defined by
\be\label{tortoise}
\frac{dr_*}{dr} = \frac{1}{f(r)} \ \ , \ \ \ \ f(r) = r^2 + \pK - \frac{2\mu}{r^{d-3}} \ee
This wave equation is to be solved for fluctuations subject to the conditions that the
flux be
ingoing at the horizon and 
outgoing at asymptotic infinity.
One thus obtains the quasinormal (QNM) modes of the black hole possessing in general a
discrete spectrum of complex frequencies whose
imaginary part determines the decay time of the small fluctuations,
\be \Im \omega = \frac{1}{\tau}\ee
The AdS/CFT correspondence maps the QNMs onto physical properties of the gauge theory fluid on the boundary. In particular, low frequency modes determine the hydrodynamic behavior of the fluid \cite{son1}, because the long-distance, low-frequency behavior of any
interacting theory at finite temperature must be described by fluid mechanics
(hydrodynamics).
This leads to a universality as
hydrodynamics implies very precise constraints on
correlation functions of conserved currents and the
stress-energy tensor.
In the hydrodynamic limit, gauge theory correlators are
fixed once a few transport coefficients are 
known. The latter may be determined by the low frequency QNMs of the dual black hole.

I shall consider various types of gravitational perturbations and calculate the corresponding low frequency QNMs following the steps in \cite{Siopsis,AS3}.

\subsection{Vector Perturbations}

The vector gravitational potential is
\be
V_V (r) =\frac{f(r)}{r^2}\left(k_V^2+\pK+\frac{(d-2)(d-4)}{4}(\pK+r^2)-\frac{3(d-2)^2\mu}{r^{d-3}}\right)
\ee
Introducing the variable
\be u=\left(\frac{r_+}{r}\right)^{d-3} \ee
the wave equation (\ref{ME}) takes the form
\be\label{waveEqn}
-(d-3)^2u^{\frac{d-4}{d-3}}\hat{f}(u)\partial_u\left(u^{\frac{d-4}{d-3}}\hat{f}(u)\partial_u \phi\right)+\hat{V}_V(u)\phi=\hat{\omega}^2\phi
\ee
where
\bes
\hat{V}_V (u)&=&\hat{f}(u)\left[\hat{k}_V^2+\frac{(d-2)(d-4)}{4}u^{\frac{2}{3-d}}-\frac{3(d-2)^2}{4}u \right. \nonumber\\
& & \left. +\frac{\pK}{r_+^2}\left(1+\frac{(d-2)(d-4)}{4}-\frac{3(d-2)^2}{4}u\right)\right]\nonumber\\
\hat{f}(u)&=& \frac{f(r)}{r^2} =1-u^{\frac{2}{d-3}}\left(u-\frac{\pK(1-u)}{r_+^2}\right)~,~~~\hat{\omega}=\frac{\omega}{r_+}~,~~~\hat{k}_V^2=\frac{k_V^2}{r_+^2}
\ees
Factoring out the behavior of the wavefunction $\phi$ at the horizon ($u=1$),
\be
\phi(u)=(1-u)^{-i\frac{\hat{\omega}}{d-1}}F(u)
\ee
the wave equation in the large $r_+$ limit (up to $\mathcal{O} (1/r_+^2)$) can be written as
\be\label{sch2} \mathcal{H} F\equiv \mathcal{A} F'' + \mathcal{B} F' + \mathcal{C} F = 0 \ee
where
\bes \mathcal{A} &=& - (d-3)^2 u^{\frac{2d-8}{d-3}} (1-u^{\frac{d-1}{d-3}}) -\pK \frac{2(d-3)^2}{r_+^2} u^2(1-u)\nonumber\\
\mathcal{B} &=& - (d-3) [ d-4-(2d-5)u^{\frac{d-1}{d-3}}]u^{\frac{d-5}{d-3}} - 2(d-3)^2 \frac{i\hat\omega}{d-1}\frac{u^{\frac{2d-8}{d-3}} (1-u^{\frac{d-1}{d-3}})}{1-u} \nonumber\\
& & - \pK \frac{d-3}{r_+^2} u\left[ (d-3)(2-3u) - (d-1) \frac{1-u}{1-u^{\frac{d-1}{d-3}}} u^{\frac{d-1}{d-3}} \right]\nonumber\\
\mathcal{C} &=& \hat{k}_V^2 + \frac{(d-2)[d-4-3(d-2)u^{\frac{d-1}{d-3}}]}{4}u^{-\frac{2}{d-3}} \nonumber\\
& &
- (d-3)\frac{i\hat\omega}{d-1} \frac{[ d-4-(2d-5)u^{\frac{d-1}{d-3}}]u^{\frac{d-5}{d-3}} }{1-u} - (d-3)^2 \frac{i\hat\omega}{d-1}\frac{u^{\frac{2d-8}{d-3}} (1-u^{\frac{d-1}{d-3}})}{(1-u)^2}\nonumber\\
& & +\pK \frac{d-2}{2r_+^2} \left[ d-4-(2d-5)u - (d-1) \frac{1-u}{1-u^{\frac{d-1}{d-3}}} u^{\frac{d-1}{d-3}} \right] \ees
Expanding the wavefunction,
\be F = F_0 + F_1 + \dots \ee
I shall solve the wave equation perturbatively.

The zeroth order wave equation,
\be\label{sch2-0} \mathcal{H}_0 F_0 = 0 \ee
is obtained in the limit $\hat\omega , \hat k_V, 1/r_+^2 \to 0$. Explicitly,
\be\label{sch20} \mathcal{H}_0 F_0= \mathcal{A}_0 F_0'' + \mathcal{B}_0 F_0' + \mathcal{C}_0 F_0 \ee
where
\bes \mathcal{A}_0 &=& - (d-3)^2 u^{\frac{2d-8}{d-3}} (1-u^{\frac{d-1}{d-3}})\nonumber\\
\mathcal{B}_0 &=& - (d-3) [ d-4-(2d-5)u^{\frac{d-1}{d-3}}]u^{\frac{d-5}{d-3}} \nonumber\\
\mathcal{C}_0 &=& \frac{(d-2)[d-4-3(d-2)u^{\frac{d-1}{d-3}}]}{4}u^{-\frac{2}{d-3}}  \ees
Two exact linearly independent solutions are
\be
F_0=u^\frac{d-2}{2(d-3)}~,~~~ \check{F}_0=u^{-\frac{d-4}{2(d-3)}}~_2F_1\left(1,-\frac{d-3}{d-1},\frac{2}{d-1};u^{\frac{d-1}{d-3}}\right)
\ee
$F_0$ is well behaved at both the horizon ($u\to1$) and the boundary ($u\to0$),
whereas
$\check F_0$ diverges at both ends making it unacceptable.

The first order equation
\be
\mathcal{H}_0 F_1+\mathcal{H}_1 F_0=0
\ee
is solved by
\be
F_1=-F_0\int \frac{\check{F}_0\mathcal{H}_1 F_0}{\mathcal{A}_0\mathcal{W}_0}+\check{F}_0\int \frac{F_0 \mathcal{H}_1 F_0}{\mathcal{A}_0 \mathcal{W}_0}
\ee
where $\mathcal{W}_0$ is the zeroth order Wronskian
\be
\mathcal{W}_0 =\frac{1}{u^{\frac{d-4}{d-3}}\left(1-u^{\frac{d-1}{d-3}}\right)}
\ee
One arrives at the constraint
\be\label{eqco} \int_0^1 \frac{F_0\mathcal{H}_1 F_0}{\mathcal{A}_0 \mathcal{W}_0} = 0 \ee
which is a linear equation in $\hat\omega$ whose solution is
\be\label{eq37}
\hat{\omega} = -i\frac{\hat{k}_V^2-\pK \frac{d-3}{r_+^2}}{d-1}
\ee
which yields the frequencies of the lowest-lying vector quasinormal modes.

For $\pK = +1$, we obtain the frequency
\be\omega = -i \frac{(l+d-2)(l-1)}{(d-1)r_+} \ee
and the maximum lifetime
\be \tau_{\mathrm{max}} = \frac{4\pi}{d}\, T_H \ee
For $\pK = 0$, the frequency is
\be \omega = -i \frac{k^2}{(d-1)r_+} \ee
leading to the diffusion constant $D = \frac{1}{4\pi T_H}$.

Finally for $\pK = -1$,
\be\label{Vsoln2}
\omega
= -i \frac{\xi^2 + \frac{(d-1)^2}{4}}{(d-1)r_+} \ \ ,
\ \ \tau = \frac{1}{|\omega|} < \frac{16\pi}{(d-1)^2}\, T_H \ee
It should be noted that in the physically relevant case $d=5$, the $K=-1$ modes live the longest, which is important for the behavior of the plasma.

\subsection{Scalar Perturbations}

In the case of scalar gravitational perturbations the potential is
\bes\label{ScalarPot}
\hat{V}_S(u)&=&\frac{u^{-\frac{2}{d-3}}-u+\frac{\pK}{r_+^2}(1-u)}{4(\hat{m}+u)^2}\nonumber\\
&\times&\Bigg\{(-6+d) (-4+d) \hat{m}^2-6 (-4+d) (-2+d) \hat{m} u+(-2+d) d u^2\nonumber\\
&-&3 (-6+d) (-2+d) \hat{m}^2 u^{\frac{d-1}{d-3}}\nonumber\\
&+&2 (18+d (-11+2d)) \hat{m} u^{\frac{2(d-2)}{d-3}}+(-2+d)^2 u^{\frac{3d-7}{d-3}}+2 (-2+d) (-1+d) \hat{m}^3 u^{\frac{2}{-3+d}}\nonumber\\
&+&\frac{\pK u^{\frac{2}{d-3}}}{r_+^2}[(-2+d) \hat{m}^2 (d+2 (-1+d) \hat{m})-3 (-2+d) \hat{m} (-8-6 \hat{m}+d (2+\hat{m})) u\nonumber\\
&+&(24+36 \hat{m}+d
(-10+d-22 \hat{m}+4 d \hat{m})) u^2+(-2+d)^2 u^3]\Bigg\}
\ees
where
\be\hat{m}=2\frac{k_S^2-\pK(d-2)}{(d-1)(d-2)(r_+^2+\pK)}
\ee
Evidently, this potential possesses a new singularity at $u =-\hat{m}$.
It is convenient to
factor out the behavior at the singularity as well as the two ends ($u=0,1$), 
\be
\phi(u)=(1-u)^{-i\frac{\hat{\omega}}{d-1}}\ \frac{u^{\frac{d-4}{2(d-3)}}}{\hat{m}+u}\ F(u)
\ee
Proceeding as with vector perturbations, after some tedious algebra \cite{Siopsis,AS3}, one arrives at the constraint
\be
\frac{d-1}{2}\, \frac{1+(d-2)\hat{m}}{(1+\hat{m})^2}+\frac{\pK}{r_+^2}\left(\frac{1}{\hat{m}}+\mathcal{O}(1)\right)-i\hat{\omega}\frac{d-3}{(1+\hat{m})^2}-\hat{\omega}^2\left(\frac{1}{\hat{m}}+\mathcal{O}(1)\right)=0
\ee
whose solution for small $\hat m$ is
\be
\hat{\omega} = \pm\sqrt{\frac{d-1}{2}\hat{m} + \frac{\pK}{r_+^2}}-i\frac{d-3}{2}\hat{m}
\ee
yielding the spectrum
\be\label{Ssoln}
\omega=\pm\frac{k_S}{\sqrt{d-2}}-i\frac{d-3}{(d-1)(d-2)r_+}\left[k_S^2-\pK(d-2)\right]
\ee
For $\pK = +1$, the frequencies are
\be\omega = \frac{l(l+d-3)}{\sqrt{d-2}} -i \frac{(d-3)(l+d-2)(l-1)}{(d-1)(d-2)r_+} \ee
and the maximum lifetime is
\be \tau_{\mathrm{max}} = \frac{d-2}{(d-3)d}\, 4\pi T_H \ee
For $\pK = 0$,
\be \omega=\pm\frac{k}{\sqrt{d-2}}-i\frac{d-3}{(d-1)(d-2)r_+}\ k^2 \ee
leading to the speed of sound $v = \frac{1}{\sqrt{d-2}}$ (which is the correct value for a CFT) and the diffusion constant $D = \frac{d-3}{d-2}\ \frac{1}{4\pi T_H}$.

For $\pK = -1$,
\be\label{Vsoln}
\omega
= \pm \sqrt{\frac{\xi^2 + (\frac{d-3}{2})^2}{d-2}} -i \frac{(d-3)[\xi^2 + \frac{(d-1)^2}{4}]}{(d-1)(d-2)r_+} \ \ ,
\ \ \tau < \frac{4(d-2)}{(d-3)(d-1)^2}\, 4\pi T_H \ee
Notice that for $d=5$, $K=-1$ scalar modes live longer than any other modes.

\subsection{Tensor Perturbations}

For tensor gravitational perturbations, the potential in the large $r_+$ limit is
\be
\hat{V}_T(u)=\frac{d-2}{4}\left(d u^{-\frac{2}{d-3}}-(d-2)u^{\frac{2(d-2)}{d-3}}-2u\right)+\hat{k}_T^2\left(1-u^{\frac{d-1}{d-3}}\right)
\ee
The zeroth order wave equation can be solved to find two independent solutions
\be
\phi_0=u^{-\frac{d-2}{2(d-3)}}~,~~~\check{\phi}_0=u^{-\frac{d-2}{2(d-3)}}\ln\left(1-u^{\frac{d-1}{d-3}}\right)
\ee
They are both unacceptable as they diverge at the boundary ($u\to 0$) and the horizon ($u\to 1$).
Therefore, no low lying tensor modes affecting the hydrodynamic behavior of the plasma exist.

\section{Hydrodynamics on the AdS boundary}
\label{sec4}

The QNMs calculated above affect the hydrodynamic behavior of the dual plasma on the AdS boundary.
This can be seen by calculating the hydrodynamics in the linearized regime of a $d-1$ dimensional fluid with dissipative effects.

The metric on the AdS boundary is
$$ds_{\partial}^2=-dt^2+d\Sigma_{\pK,d-2}^2$$
The hydrodynamic equations amount to the conservation law of the stress-energy tensor,
\be \nabla_\mu T^{\mu\nu}=0
\ee
The stress-energy tensor can be cast in the general form
\be\label{Stensor}
T^{\mu\nu} = (\epsilon+p)u^\mu u^\nu+p g^{\mu\nu}-\eta\left(\triangle^{\mu\lambda}\nabla_\lambda u^\nu+\triangle^{\nu\lambda}\nabla_\lambda u^\mu-\frac{2}{d-2}\triangle^{\mu\nu}\nabla_\lambda u^\lambda\right)-\zeta\triangle^{\mu\nu}\nabla_\lambda u^\lambda
\ee
where $\triangle_{\mu\nu}=g_{\mu\nu}+u_\mu u_\nu$ and $\epsilon$, $p$, $\eta$ and $\zeta$ represent the energy density, pressure, shear viscosity and bulk viscosity, respectively, of the gauge theory fluid.
Two constraints on the parameters immediately follow from the requirement of conformal invariance,
\be\label{eqpz} \epsilon=(d-2)p\ \ , \ \ \ \ \zeta=0\ee
$u^\mu$ is the velocity field of the fluid.  In its rest frame, $u^\mu=(1,0,0,0)$ and the pressure is constant, $p=p_0$. Perturbations introduce small disturbances,
\be
u^\mu=(1,u^i)
\ee
where $u^i$ is small and also allow for small corrections to the pressure so that
\be p=p_0+\delta p\ee
The hydrodynamic equations imply
\bes
(d-2)\partial_t \delta p+(d-1)p_0 \nabla_i u^i&=&0\nonumber\\
(d-1)p_0 \partial_t u^i +\partial^i\delta p-\eta\left[\nabla^j \nabla_j u^i+\pK(d-3) u^i+\frac{d-4}{d-2}\partial^i (\nabla_j u^j)\right]&=&0
\ees
where I used $R_{ij}=\pK(d-3)g_{ij}$.

For vector perturbations, use the {\it ansatz}
\be
\delta p=0~,~~~u^i=\mathcal{C}_V e^{-i\omega t}{\mathbb V}^i
\ee
where ${\mathbb V}^i$ is a vector harmonic.

The hydrodynamic equations become
\be\label{Vhydro}
-i \omega (d-1) p_0+\eta\left[k_V^2-\pK(d-3)\right]=0
\ee
Using
\be\label{eq62}
\frac{\eta}{p_0}=(d-2)\frac{\eta}{s}\frac{S}{M}=\frac{4\pi\eta}{s}\frac{r_+}{\pK+r_+^2}
\ee
with $\omega$ from the gravity dual (eq.~(\ref{eq37})), we obtain
for large $r_+$,
\be\label{Shydro}
\frac{\eta}{s} =\frac{1}{4\pi}
\ee
as expected \cite{son1}.

For scalar perturbations, use instead the {\it ansatz}
\be
u^i=\mathcal{A}_S e^{-i\omega t}\partial^i \mathbb{S}~,~~~\delta p=\mathcal{B}_S e^{-i\omega t}\mathbb{S}
\ee
where $\mathbb{S}$ is a scalar harmonic.

The hydrodynamic equations in this case read
\bes
&&(d-2)i\omega \mathcal{B}_S+(d-1)p_0 k_S^2\mathcal{A}_S=0\nonumber\\
&&\mathcal{B}_S+\mathcal{A}_S\left[-i\omega(d-1)p_0-2(d-3)\pK\eta +2 \eta k_S^2\frac{d-3}{d-2} \right]=0
\ees
This is a linear system whose determinant must vanish,
\be \left| \begin{array}{cc} (d-2)i\omega & (d-1) p_0 k_S^2 \\
1 & -i\omega (d-1) p_0 - 2(d-3)\pK\eta + 2\eta k_S^2 \,\frac{d-3}{d-2} \end{array} \right| = 0 \ee
Using this constraint and arguing along the same lines as with vector perturbations (eq.~(\ref{eq62})), one arrives at
the same special value of the ratio $\eta/s$ (\ref{Shydro}), as before.


The fluid described above is static with small sinusoidal perturbations. This is hardly the behavior of the quark-gluon plasma observed at RHIC. However, it was argued in \cite{FGMP} that a flowing behavior can be obtained in the spherical case ($\pK =+1$) by choosing a foliation near the AdS boundary which consists of flat Minkowksi hypersurfaces instead of ones consisting of spatial spheres ($S^{d-2}\times \mathbb{R}$).
The latter are obtained by fixing the radial distance in the metric ($r=$~const.~hypersurfaces).
The former (which are physically relevant) are obtained from the latter via a conformal transformation.
The static fluid is then mapped onto a ``conformal soliton flow'' whilst the perturbations induce an elliptic flow similar to the one observed at RHIC.
One may then calculate physical quantities and compare with experimental results.

Let $v_2 = \langle \cos 2\phi \rangle$ at $\theta = \frac{\pi}{2}$ (mid-rapidity) averaged with respect to the energy density at late times and
$\delta = \frac{\langle y^2 - x^2 \rangle}{\langle y^2 + x^2 \rangle}$ (eccentricity at time $t=0$).
The ratio of these two quantities is found to be \cite{FGMP}
\be \frac{v_2}{\delta} = \frac{1}{6\pi} \Re \frac{\omega^4-40\o^2+72}{\o^3-4\o} \sin \frac{\pi\o}{2} \ee
where $\omega$ is the frequency of the longest lived (scalar) QNM.
Numerically, $\frac{v_2}{\delta} = 0.37$, which compares well with the result from RHIC data, $\frac{v_2}{\delta} \approx 0.323$
\cite{bibPHENIX}.

Another quantity that can be predicted is the thermalization time,
\be \tau = \frac{1}{2|\Im \omega|} \approx \frac{1}{8.6 T_{\mathrm{peak}}} \approx 0.08~\mathrm{fm/c}\ \ , \ \ \ \ T_{\mathrm{peak}} = 300~\mathrm{MeV} \ee
to be compared with the RHIC result $\tau \sim 0.6$~fm/c \cite{bibALMY}.
While not in good agreement, it is encouragingly small.
For comparison, 
perturbative QCD yields $\tau \gtrsim 2.5$~fm/c \cite{bibBMSS,bibMG}.

A similar comparison with experimental results can also be done in the case of a hyperbolic horizon (topological black hole, $\pK =-1$).
Recall that this is the case of longest lived QNMs, so these modes may play an important role in the late time behavior of the quark-gluon plasma.
However, it is much harder than with spherical horizons to extract predictions
because of the complexity of the 
conformal map $\mathbb{H}^{d-2}/\Gamma \times \mathbb{R} \mapsto (d-1)-$dim Minkowski space \cite{AS3}.


\section{Bjorken flow}
\label{sec5}

As was first suggested by Bjorken \cite{Bjorken}, in high energy collisions of heavy nuclei,
there is a
``plateau'' in particle production in the central rapidity region
since the interactions are independent of the Lorentz frame (boost-invariant)
and the emergent nuclei are highly Lorentz-contracted pancakes
receding almost at the speed of light.

The initial conditions are shown in table \ref{tablex} for RHIC and the LHC.
$\tau$ is the longitudinal proper time, $\epsilon$ is the energy density, $T$ is the temperature, and $\sqrt s$ is the center of mass energy of the beams.

\begin{table}[h]
\caption{\label{tablex}Initial conditions.}
\begin{center}
\begin{tabular}{||l||c|c|c|c||}
\hline\hline
 & $\tau_0$ (fm/c) & $\epsilon_0$ (GeV/fm$^3$) & $T$ (GeV) & $\sqrt s$ (GeV) \\
\hline
\hline
RHIC &  0.2 & 10 & 0.5 & 200 \\
\hline
LHC &  0.1 & 10 &  1 & 5,500 \\
\hline
\hline
\end{tabular}
\end{center}
\end{table}
The hydrodynamic equations
respect the symmetry of the initial conditions (boost invariance)
and lead to simple solutions.
For a conformal fluid,
\be \epsilon \sim \tau^{-4/3} \ \ , \ \ \ \ T \sim \tau^{-1/3} \ \ , \ \ \ \ s \sim \tau^{-1} \ee
To understand Bjorken Hydrodynamics, consider a
plasma on a $(d-1)$-dimensional Minkowski space spanned by
coordinates $\tilde x^\mu$ ($\mu = 0,1,\dots, d-2$) with the colliding beams along the $\tilde x^1$ direction.

Define coordinates $\tau, y$ (proper time and rapidity $y$ in the longitudinal plane, respectively) by
\be
\tilde x^0=\tau \cosh y~~,~~~~\tilde x^1=\tau \sinh y
\ee
The Minkowski metric reads
\be\label{eq16}
d\tilde s^2=d\tilde x_\mu d\tilde x^\mu = -d\tau^2+\tau^2 dy^2+(d\tilde x^\bot)^2
\ee
where $\tilde x^\bot = (\tilde x^2,\dots,\tilde x^{d-2})$ are the transverse coordinates.

Assuming no viscosity, the stress-energy tensor is
\be
T^{\mu\nu} =
\left(\begin{array}{cccc}
\epsilon (\t) & 0 & \dots & 0 \\
0 & p(\t)/\t^2 & \dots & 0\\
~ & ~ & \ddots & ~\\
0 & 0 & \dots & p(\t)
\end{array}\right) \ .
\ee
where $\epsilon$ and $p$ are the energy density and pressure of the fluid, respectively.
The local conservation law of the stress-energy tensor implies
\be
\p_\t \epsilon +\frac{1}{\t} (\epsilon + p)  =0
\ee
Conformal invariance implies tracelessness,
$
-\epsilon +(d-2)p=0
$,
%
therefore
\be\label{HydroEP}
\epsilon = (d-2) p = \frac{\epsilon_0}{\tau^{\frac{d-1}{d-2}}}
\ee
Then conservation of entropy in a perfect fluid implies
\be\label{HydroT}
T=\frac{T_0}{\tau^{1/(d-2)}}
\ \ , \ \ \
s = \frac{\dot p}{\dot T} = \frac{s_0}{\tau} \ \ , \ \ \ \ s_0 = \frac{d-1}{d-2}\, \frac{\epsilon_0}{T_0} \ee
Note that the energy and entropy densities have the same dependence on the temperature as in static case (discussed in section \ref{sec2} in terms of a gravity dual).

A gravity dual of the above perfect fluid flow was obtained by Janik and Peschanski \cite{JP}.
They obtained an asymptotic solution of the Einstein equations
which was
valid for large longitudinal proper time $\tau$.
Their metric reads
\be\label{metricbj}
ds^2_{\mathrm{Bjorken}} =\frac{1}{\tilde z^2} \left[ -\left( 1 -2\mu \frac{\tilde z^{d-1}}{\tau^{\frac{d-1}{d-2}}} \right) d\tau^2+\tau^2 dy^2+(d\tilde x^\bot)^2
+\frac{d\tilde z^2}{ 1 -2\mu \frac{\tilde z^{d-1}}{ \tau^{\frac{d-1}{d-2}} } }  \right]
\ee
and leads to Bjorken hydrodynamics via holographic renormalization.


In \cite{AS} it was shown that Bjorken hydrodynamics can be obtained from an AdS Schwarzschild black hole. To see this,
instead of
approximating the boundary of the flat AdS black hole (\ref{eqmebh}) with $z=$~const.~hypersurfaces (as $z\to 0$), slice with
$\tilde z=$~const.~hypersurfaces, where
\be\label{eq77}
t=\frac{d-2}{d-3} \tau^{\frac{d-3}{d-2}}~~,~~~~x^1=\tau^{\frac{d-3}{d-2}} y
~~,~~~~x^\bot=\frac{\tilde x^\bot}{\tau^{1/(d-2)}}~~,~~~~z=\frac{\tilde z}{\tau^{1/(d-2)}}
\ee
The two different foliations
coincide ``initially'' (at $\tau = 1$ in arbitrary units; see table \ref{tablex} for conventional units).
Let us identify initial data with their corresponding values in the static case (eqs.~(\ref{eqTH}) and (\ref{eq15})),
\be\label{eqTe} T_0 = T_H \ \ , \ \ \ \ \epsilon_0 = \frac{(d-2)\mu}{8\pi G} \ee
We obtain a flowing fluid on the boundary because the new coordinates (\ref{eq77}) of the black hole metric (\ref{eqmebh}) are $\tau$-dependent.

Applying the transformation (\ref{eq77}) to the exact black hole metric (\ref{eqmebh}), or
more precisely, to a patch which includes the boundary $z\to 0$,
the black hole metric becomes
\be\label{TransMetric}
ds^2_{\mathrm{b.h.}} = \frac{1}{\tilde z^2} \left[ - \left( 1-2\mu \frac{\tilde z^{d-1}}{\tau^{\frac{d-1}{d-2}}} \right) d\tau^2 + \tau^2 dy^2 + (d\tilde x^\bot)^2
+ \frac{d\tilde z^2}{1-2\mu \frac{\tilde z^{d-1}}{\tau^{(d-1)/(d-2)}}} \right] + \dots
\ee
where the dots represent higher-order contributions in the large $\tau$ expansion.
This form matches the bulk metric of Bjorken flow (\ref{metricbj}) to leading order in $1/\t$.

Having a dual black hole also allows us to understand the origin of the temperature of the quark-gluon plasma.
Recall that the Hawking temperature of the hole $T_H$ (\ref{eqTH}) is the temperature of the {\it static} fluid on the $z=0$ hypersurface whose metric is
\be\label{eqm1} ds_{z \to 0}^2 = - dt^2 + d\vec x^2 \ee
On the other hand, the $\tilde z \to 0$ hypersurface has metric
\be\label{eqm2} ds_{\tilde z \to 0}^2 = -d\t^2 + \t^2 dy^2 + (d\tilde x^\bot)^2 \ee
The two metrics are related through the transformation
\be\label{eq34b}
t=\frac{d-2}{d-3} \tau^{\frac{d-3}{d-2}}~~,~~~~x^1=\tau^{\frac{d-3}{d-2}} y ~~,~~~~
x^\bot=\frac{\tilde x^\bot}{\tau^{1/(d-2)}}
\ee
by a leading-order conformal factor,
\be
ds_{z \to 0}^2 = \tau^{-\frac{2}{d-2}}\left[ ds_{\tilde z \to 0}^2 + \dots \right]
\ee
It follows that the period of thermal Green functions
on the Bjorken boundary scales as $\tau^{1/(d-2)}$.
Since the period is inversely proportional to the temperature, the latter scales as $\tau^{-1/(d-2)}$, in agreement with Bjorken hydrodynamics.
The two hypersurfaces coincide at $\t = 1$ at which time $T=T_H$.

The above asymptotic expressions are exact (no higher order corrections) in three dimensions ($d=3$) as was discovered in \cite{KLT}.
Indeed, letting $d\to 3$, the coordinate transformation (\ref{eq77}) becomes
\be
t=\ln~\tau~~,~~~~x^1 =y~~,~~~~z=\frac{\tilde z}{\tau}
\ee
and the transformation to Fefferman-Graham coordinates can be found exactly in this case,
\be z = \frac{\tilde z_{FG}}{\tau} \left( 1 + \frac{\mu}{2}\, \frac{\tilde z_{FG}^2}{\tau^2} \right)^{-1} \ee
leading to explicit expressions (in closed form) for the physical properties of the plasma.

Higher order corrections to the Bjorken flow dictated by the black hole may be found by refining the transformation (\ref{eq77}).
At next-to-leading order, viscosity is introduced whereas at higher orders
it is no longer possible to maintain a metric which is independent of the rapidity and the transverse coordinates.
For Bjorken flow, it is necessary to introduce a perturbation.
The resulting metric leads to a fluid on which the ratio $\eta/s$ takes on the special value (\ref{eq1}) \cite{AS2} in agreement with sinusoidal perturbations \cite{son1}.

%
%

\section{Conclusion}
\label{sec6}

Quasi-normal modes are a powerful tool in understanding the hydrodynamic behavior of the gauge theory fluid at strong coupling
via the AdS/CFT correspondence.
I analytically calculated the low-lying quasinormal modes of AdS black holes in arbitrary dimension in the high temperature limit.
I considered all three different types of perturbations (scalar, vector and tensor) and solved the wave equation \cite{IK} in each case by applying the methods in~\cite{Siopsis,AS3}.
I obtained quasinormal frequencies which were in agreement with the frequencies obtained by considering perturbations of the gauge theory fluid on the boundary \cite{FGMP,MP}.
These modes may contribute to the properties of the quark-gluon plasma produced in heavy ion collisions.

I also discussed the possibility of obtaining viscous Bjorken hydrodynamics on a $(d-1)$-dimensional Miskowski space from a large AdS$_d$ Schwarzschild black hole (of flat horizon).
The latter is normally considered dual to a static gauge theory fluid on the boudary whose temperature coincides with the Hawking temperature.
By appropriately modifying the boundary conditions, I obtained Bjorken hydrodynamics on the boundary in the limit of large longitudinal proper time \cite{AS,AS2}. These results are in agreement with those obtained by considering time-dependent asymptotic solutions of the Einstein equations in five dimensions~\cite{JP,NakSin,SinNakKim}.
Moreover, since the bulk space consists of a Schwarzschild black hole,
I could determine the temperature of the conformal fluid on the boundary in terms of the Hawking temperature of the hole.
It turns out that at next-to-next-to-leading order, no choice of boundary conditions leads to a boost-invariant flow \cite{AS2}. In order to obtain a dual Bjorken flow at that order, the black hole metric needs to be perturbed. This leads to a constraint on the viscosity coefficient and the viscosity to entropy density ratio is then fixed to the value $1/(4\pi)$ as in the case of sinusoidal perturbations \cite{son1,son2,son3}. This is also in agreement with a next-to-next-to-leading order calculation of a time-dependent asymptotic solution \cite{Janik}.

It should be pointed out that a constraint on the viscosity is not necessary if one does not perturb the black hole metric. In this case, one obtains deviations from Bjorken flow which are a subleading effect at late times. It might be worth exploring the connection of such deviations (coming from a dual Schwarzschild black hole) to experimental data (RHIC and the LHC).


In conclusion, it is an exciting possibility that heavy ion collisions at RHIC and the LHC will probe black holes and provide information on string theory
as well as non-perturbative QCD effects.



\end{document}